\documentclass[aps,prb,twocolumn, superscriptaddress,nobibnotes]{revtex4-1}

\usepackage{graphicx}
\usepackage{pstricks}
\usepackage{amsmath}
\usepackage{gensymb}
\usepackage{amsfonts}
\usepackage{dcolumn}
\usepackage{bm}

\newcommand{\bra}[1]{\langle#1|}
\newcommand{\ket}[1]{|#1\rangle}

\begin{document}
	
	\title{Superconductivity in correlated BEDT-TTF
		molecular conductors: critical temperatures and gap symmetries}
	
	\author{Karim Zantout}
	\email{zantout@itp.uni-frankfurt.de}
	\affiliation{Institut f\"ur Theoretische Physik, Goethe-Universit\"at 
		Frankfurt, 
		Max-von-Laue-Stra{\ss}e 1, 60438 Frankfurt am Main, Germany}		
	
	\author{Michaela Altmeyer}
	\affiliation{Institut f\"ur Theoretische Physik, Goethe-Universit\"at 
		Frankfurt, 
		Max-von-Laue-Stra{\ss}e 1, 60438 Frankfurt am Main, Germany}
	
	\author{Steffen Backes}
	\email{present address: Centre de Physique Th\'eoretique, \'Ecole Polytechnique, F-91128 Palaiseau, France}
	\affiliation{Institut f\"ur Theoretische Physik, Goethe-Universit\"at 
		Frankfurt, 
		Max-von-Laue-Stra{\ss}e 1, 60438 Frankfurt am Main, Germany}
	\author{Roser Valent\'i}
	\affiliation{Institut f\"ur Theoretische Physik, Goethe-Universit\"at 
		Frankfurt, 
		Max-von-Laue-Stra{\ss}e 1, 60438 Frankfurt am Main, Germany}
	
	\begin{abstract}
		Starting from an {\it ab initio}-derived two-site dimer Hubbard hamiltonian
		on a triangular lattice,
		we calculate  the superconducting gap functions and  critical
		temperatures for representative $\kappa$-(BEDT-TTF)$_2$X superconductors
		by solving the linearized Eliashberg equation using the Two-Particle Self-Consistent
		approach (TPSC)  extended to multi-site problems. Such an extension allows
		for the inclusion of molecule degrees of freedom in the description of
		these systems. We present
		both, benchmarking results for the half-filled dimer model as well as detailed
		investigations for the 3/4-filled molecule model. 
		Remarkably,  we find in the latter model that the phase
		boundary between the two most competing gap symmetries discussed
		in the context of these materials -- d$_{xy}$ and the
		recently proposed eight-node s+d$_{x^2-y^2}$ gap symmetry -- 
		is located within the regime of
		realistic model parameters  and is especially sensitive to the degree of
		in-plane anisotropy in the materials
		as well as to the value of the on-site Hubbard repulsion. We show that
		these results provide a more complete and accurate  description
		of the superconducting properties of $\kappa$-(BEDT-TTF)$_2$X 
		than previous Random Phase Approximation (RPA) calculations and, in particular, we
		discuss predicted critical temperatures in comparison to experiments. Finally, our  
		findings suggest that it may be even easier to experimentally switch
		between the two pairing symmetries as previously anticipated by invoking
		pressure, chemical doping or disorder effects.
	\end{abstract}

	
	\maketitle
	
	\section{Introduction}
	Among the classes of quasi two-dimensional organic charge transfer salts, the
	$\kappa$-(BEDT-TTF)$_2$X family, often abbreviated as $\kappa$-(ET)$_2$X, is of
	special interest since its members exhibit rich phase diagrams with
	antiferromagnetic Mott insulating, superconducting (SC), and spin-liquid
	states~\cite{Toyota2007, Powell2005, Shimizu2003}. Besides chemical
	substitution of the monovalent anion $X^-$ and/or physical
	pressure~\cite{Dumm2009,Kawaga2005} the $\kappa$-(BEDT-TTF)$_2$X salts offer the possibility
	to tune between the different states by endgroup disorder
	freezing~\cite{Toyota2007,Hartmann2014,Guterding2015}.
	
	Measurements of electronic properties such as  specific heat, conductivity
	or magnetic susceptibility~\cite{Toyota2007} 
	evidence a strong anisotropy between the stacking direction and the
	two-dimensional ET-planes, which may even become superconducting below
	transition temperatures of about
	$10$~K~\cite{Hiramatsu2015,Kato1987,Mori1990,Kini1990}.  Even though a large
	variety  of experimental techniques has been employed to study the character of
	the superconducting order parameter, no consensus on the symmetry of the gap
	function has been reached so far and proposals range from
	$s$-wave~\cite{Elsinger2000,Mueller2002,Wosnitza2003} to
	$d$-wave~\cite{Taylor2007,Taylor2008,Malone2010,Milbradt2013,Izawa2001,Schrama1999,Arai2001,Ichimura2008,Oka2015}
	states. Even within the group of researchers that agree on a $d$-wave
	superconducting order parameter, there are controversial measurements regarding
	the position of the nodes on the Fermi 
	surface~\cite{Izawa2001, Malone2010,Diehl2016,Kuehlmorgen2017}. 
	However, the similar phase diagrams (antiferromagnetic Mott and SC phase) of
	the high-temperature cuprate superconductors and $\kappa$-(ET)$_2$X suggest a
	common pairing mechanism based on antiferromagnetic spin fluctuations, although
	the additional effect of geometrical frustration in the $\kappa$-(ET)$_2$X
	family yields another degree of complexity with not yet completely understood
	consequences~\cite{McKenzie1997,Kanoda_review2017}.
	
	In a recent study\cite{Guterding2016,Diehl2016}, a comparison of the
	widely used dimer model and the more accurate molecule model has provided
	evidence that a strong degree of dimerization,
	characterized by the
	intra-dimer hopping, is not sufficient to guarantee the
	validity of the dimer approximation.   In contrast, it was shown that due to the in-plane anisotropy
	of the hopping parameters this approximation is not applicable to the whole
	$\kappa$-(ET)$_2$X  family, where the more accurate description within the molecule
	model even results in a different gap symmetry. 
	All materials were found to be located in the  eight-node gap s+d$_{x^2-y^2}$ region 
	of the phase diagram with some compounds close to the phase boundary to a d$_{xy}$ symmetry.
	Scanning tunneling spectroscopy measurements for $\kappa$-(ET)$_2$Cu[N(CN)$_2$]Br showed
	compatibility with the proposed eight-node gap symmetry~\cite{Diehl2016}. However, as these measurement 
can only access the absolute value of
the gap function, phase sensitive measurements will be required to 
uniquely settle this discussion also for the other members of the 
$\kappa$-ET family. 
	
	All the previously mentioned analyses were based on weak-coupling RPA calculations.
	Since providing an accurate  location of the boundary between the two gap symmetries
	may help to unveil the origin of apparent contradicting experimental observations in the $\kappa$-(ET)$_2$X  family, 
	in the present work we go beyond RPA and reanalyze the superconducting properties in these materials.
	We employ the linearized Eliashberg theory combined with an extension of the single-band intermediate-coupling
	Two-Particle Self-Consistent (TPSC) approach introduced by Vilk and Tremblay~\cite{Vilk1997}. This enables us to not only calculate the gap symmetries in the dimer and molecule model but also, in general, to determine the critical temperatures associated with the different models and materials.
	We find that in the dimer model the critical temperatures show an approximately linear dependence on the frustration ratio with d$_{xy}$ symmetry of the order parameter in agreement to previous studies~\cite{Kuroki2002, Schmalian1998}. 
	In contrast, the real $\kappa$-(ET)$_2$X  compounds do not follow this simple 
	relation further evidencing the inadequacy of the dimer approximation.
	In the molecule model we find that the inclusion of the TPSC self-energy gives
	rise to pseudogap physics preventing the transition to the superconducting
	state.  
	Although this hampers the calculation of critical temperatures, we can
	determine the gap symmetries by carefully approaching the SC phase and find that
	besides a large in-plane anisotropy and large intra-dimer hopping, strong correlations stabilize gap functions with extended s+d$_{x^2-y^2}$ symmetry at low temperatures.
	Hence, the gap symmetry in these materials is determined by the complex interplay of these experimentally highly tunable parameters.
	Based on these results we conclude that small changes on the crystal structure introduced via pressure, chemical doping, or disorder may easily switch between the different symmetry states.

	\section{Methods and Models}
	\subsection{Ab-initio calculations and model Hamiltonian}
	\label{sec:metfhodsabinitio}
	As the electronic properties of the $\kappa$-(ET)$_2$X systems, such as  
	the conductivities, are highly
	anisotropic with the largest contribution within the ET planes, it is justified
	to focus only on these two-dimensional planes. The $\kappa$ packing motif
	allows for two distinct model descriptions with different degrees of
	approximation. The dimer model constitutes the strongest simplification, in
	which the center of two parallel ET molecules is taken as a single lattice
	site, resulting in a half-filled anisotropic triangular lattice model
	with two dimers in the crystallographic unit cell~\cite{Guterding2016}
	(two-band model).
	The molecular model further resolves the inner structure of each dimer as each
	individual molecule corresponds to a tight binding lattice site yielding a
	three quarter filled  model with four molecules per crystallographic
	unit cell (four-band model).
	
\begin{figure}[t]
		\includegraphics[width=0.8\linewidth]{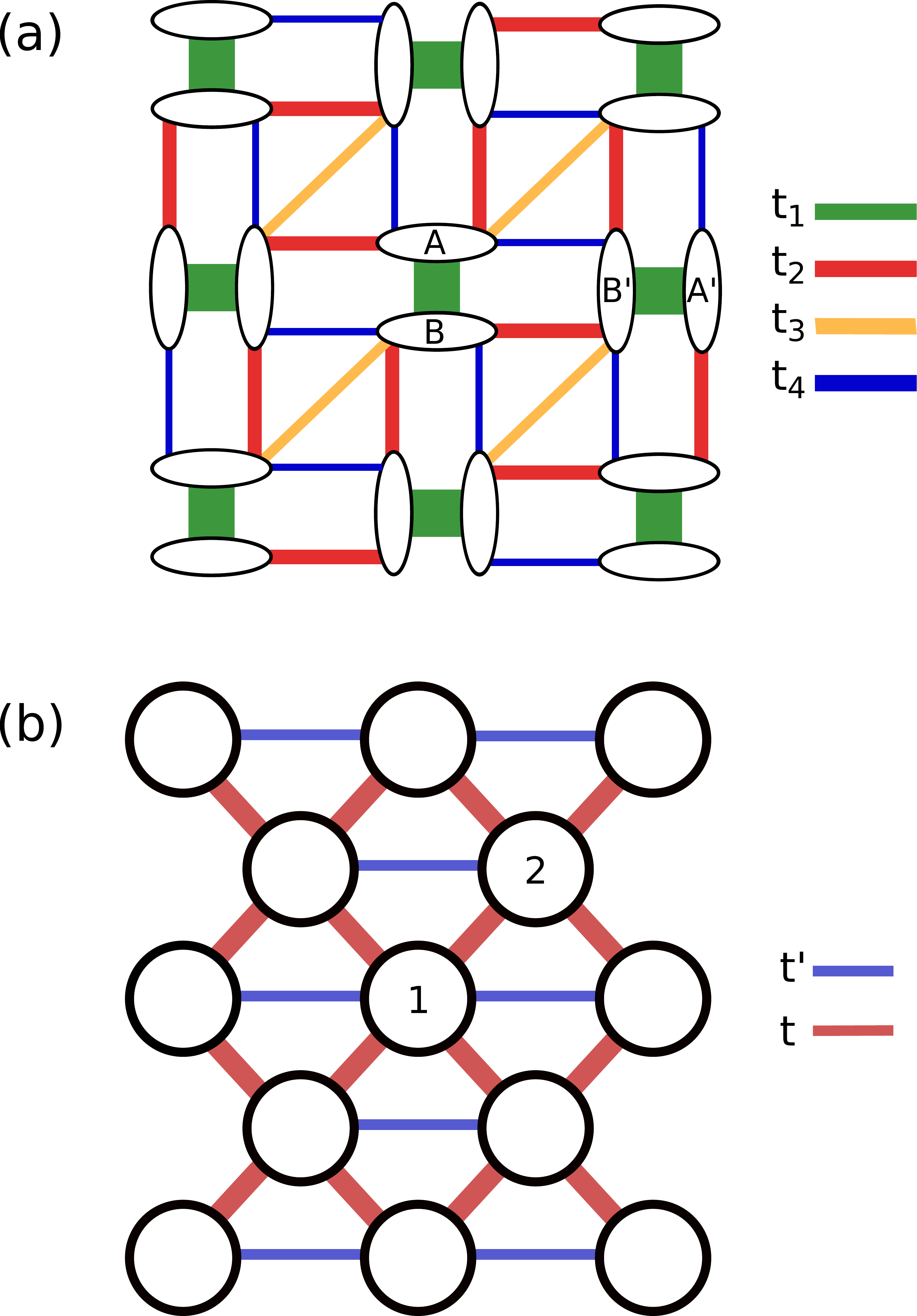}
		\caption{(a) Lattice structure within the molecule model of the $\kappa$-(ET)$_2^+$ layer. 
			ET molecules are single lattice sites (ellipsoids in figure). The amplitude of the
			four dominant hopping integrals ($t_1$,$t_2$,$t_3$,$t_4$) are shown by the thickness of the line. (b) In the dimer model one has to integrate out $t_1$ and average between $t_2$ and $t_4$ that results in the dominant hopping integral $t$ while $t'$ is connected to the former $t_3$ by dividing contributions from the two separate ET molecules that are counted now as one.}
		\label{fig:lattice}
	\end{figure}

	Using the projective Wannier function method as implemented in FPLO, the hopping
	amplitudes between the localized molecular orbitals have been
	calculated~\cite{Guterding2016}, where the large differences in the order of
	magnitude allow us to neglect all but four hopping parameters
	($t_1,t_2,t_3,t_4$, see Fig.~\ref{fig:lattice}) in a first approximation since the next order of hopping elements is about 10\% of the smallest hopping $t_4$.  Our
	kinetic energy Hamiltonian is then given by
	\begin{align}
	H_\text{kin} &= \sum_{ij,\alpha,\beta,\sigma} t_{ij}^{\alpha\beta}(c^\dagger_{\beta \sigma}(\vec r_j) c^{\,}_{\alpha \sigma}(\vec r_i) + h.c.) \nonumber\\
	&-\mu \sum_{i,\alpha,\sigma}c^\dagger_{\alpha,\sigma}(\vec r_{i}) c_{\alpha,\sigma}(\vec r_{i})\label{eq:kin_hamiltonian_4} 
	\end{align}
	where $t_{ij,\alpha\beta}$ are the hoppings from site $\alpha$ in unit cell $i$ to site $\beta$ in unit cell $j$, $\mu$ is the chemical potential, $c^\dagger_{\alpha \sigma}(\vec r_i)$ creates an electron in unit cell $i$ at site $\alpha$ with spin $\sigma$ while $c^{\,}_{\alpha \sigma}(\vec r_i)$ annihilates an electron in unit cell $i$ at site $\alpha$ with spin $\sigma$.
	
	The parameters of the dimer model
	\begin{align}
	H_\text{kin} &= \sum_{\langle ij\rangle,\alpha,\sigma } t\left(c^\dagger_{\alpha,\sigma}(\vec r_i) c_{\alpha,\sigma}(\vec r_{j}) + h.c.\right)\nonumber\\
	&+\sum_{[ij],\alpha,\beta,\alpha\neq\beta } t'\left(c^\dagger_{\alpha,\sigma}(\vec r_{i}) c_{\beta,\sigma}(\vec r_{j}) + h.c. \right)\nonumber\\
	&-\mu \sum_{i,\alpha,\sigma}c^\dagger_{\alpha,\sigma}(\vec r_{i}) c_{\alpha,\sigma}(\vec r_{i})\label{eq:kin_hamiltonian_2} 
	\end{align}
	can be derived using the geometrical relations
	\begin{subequations}
		\begin{align}
		t & = (|t_2|+|t_4|)/2 \\
		t^\prime &= |t_3|/2.
		\end{align}
		\label{eq:geometricformulas}
	\end{subequations}
	
	The tight binding dispersions 
	can be easily determined analytically through the 
	four matrix elements between the two dimer states $\ket{\alpha=0}$
	and $\ket{\alpha=1}$ in the dimer model, 
	\begin{subequations}
		\begin{align}
		\bra{0}H_\text{kin}\ket{0}(\vec k) =& \bra{1}H_\text{kin}\ket{1}(\vec k) = 2 t^\prime \text{cos} 
		(k_x a)-\mu\\
		\bra{0}H_\text{kin}\ket{1}(\vec k) =& t\left( e^{i k_x a/2 + i k_y b/2} + e^{-i k_x a/2 + i k_y b/2} + \right.\nonumber\\
		&\left.+e^{i k_x a/2 - i k_y b/2} + e^{-i k_x a/2 - i k_y b/2}\right)\nonumber\\
		=& \bra{1}H_\text{kin}\ket{0}^\ast(\vec k)
		\end{align}
		\label{eq:twobanddimerhamil}
	\end{subequations}
	and six distinct contributions between the four
	molecule states for the molecule model
	\begin{subequations}
		\begin{align}
		\bra{0}H_\text{kin}\ket{1}(\vec k) &= t_1 + t_3 \, e^{i k_x a} \\
		\bra{0}H_\text{kin}\ket{2}(\vec k) &= t_4 \left( 1 + e^{-i k_y b} \right) \\
		\bra{0}H_\text{kin}\ket{3}(\vec k) &= t_2 \left( 1 + e^{-i k_x a} \right)  \\
		\bra{1}H_\text{kin}\ket{2}(\vec k) &= t_2 \, e^{-i k_y b } \left( 1 + e^{-i k_x a} 
		\right)  \\
		\bra{1}H_\text{kin}\ket{3}(\vec k) &= t_4 \, e^{-ik_x a} \left( 1 + e^{-i k_y b} \right) 
		\\
		\bra{2}H_\text{kin}\ket{3}(\vec k) &= t_1 + t_3 \, e^{-i k_x a}\\
		\bra{\alpha}H_\text{kin}\ket{\alpha}(\vec k) &= -\mu,
		\end{align}
		\label{eq:fourbandmoleculehamil}
	\end{subequations}
	where $a$ and $b$ are lattice constants of the two-dimensional ET plane and the remaining matrix elements are obtained from $H=H^\dag$.
	Note that the chemical potential was determined numerically to ensure the correct filling in both models.
	
	\subsection{Two-Particle Self-Consistent calculations}
	Due to the similarity of the phase diagram of cuprates and $\kappa$-(ET)$_2$ materials we can assume a spin-fluctuation based mechanism for the superconductivity. We will use an extension of the Two-Particle Self-Consistent (TPSC) approach as introduced by Vilk and Tremblay~\cite{Vilk1997} to find an approximate solution for our multi-site dimer and molecule models with Hubbard on-site interaction,
	\begin{equation}
	\begin{array}{rl}
	H =& H_\text{kin} + H_\text{int} \\ =& H_\text{kin} + \frac{U}{2} \sum\limits_{i,\alpha,\sigma} 
	n_{\alpha\sigma}(\vec r_i) n_{\alpha\bar\sigma}(\vec r_i) ,
	\end{array}
	\label{eq:hamiltonian}
	\end{equation}
	where $U$ is the Hubbard on-site interaction and $n_{i\alpha\sigma}$ is the number operator for electrons in unit cell $i$ at site $\alpha$ with spin $\sigma$.
	Note that the on-site $U$ term in the dimer model corresponds to the
	Coulomb interaction in the dimer where one can approximate~\cite{Powell2005,Powell2006} $U_\text{dim}\approx 2t_1$, while the on-site $U$ in the molecule
	model corresponds to the Coulomb interaction in the molecule $U_\text{mol}$. In the
	present work we do not include inter-site Coulomb contributions~\cite{Kaneko2017}.
Please note that the inclusion of intermolecular Coulomb repulsion 
allows for the possibility of describing charge density wave phases 
in proximity to superconductivity~\cite{Sekine2013,Watanabe2017}.
However, the observed solutions are very similar and we can only speculate that this may be due to a robustness of the instabilities 
that yield the resulting gap symmetries against further-neighbor 
interactions.
	
	So far, TPSC has been successfully applied, f.i., to investigate pseudogap physics~\cite{Tremblay2011} and the dome-like shape of the superconducting critical temperature~\cite{Ogura2015,Kyung2003} for the Hubbard model on a square lattice.
	TPSC is a conserving and self-consistent approximation, in which higher order contributions to the four-point vertices are reduced to their averages.
	As a consequence, the resulting equations yield a weak- to intermediate-coupling approach for the solution of the Hubbard model.
	
	We define the non-interacting multi-site Green's function
	\begin{equation}
	\begin{array}{rl}
	G^0_{\mu\nu}(\vec k, i\omega_n) = \left[i\omega_n\mathbb{I} - H_\text{kin}(\vec k)\right]^{-1}_{\mu\nu},
	\end{array}
	\label{eq:nonintgreen}
	\end{equation}
	where $\mu,\nu$ are site indices, $\vec k$ is a two-dimensional reciprocal lattice vector, and $\omega_n= (2n+1)\pi~T$ are fermionic Matsubara frequencies at temperature $T$.
	Moreover, we calculate the non-interacting susceptibility $\chi^0$
	\begin{equation}
	\begin{array}{rl}
	\chi^0_{\lambda\mu\nu\xi}(\vec q, iq_m) = -\frac{1}{N_{\vec k}}\sum\limits_{\vec k, b,c }&a^\nu_{b}(\vec k)a^{\lambda\ast}_b(\vec k)a^{\mu}_c(\vec k+\vec q)a^{\xi\ast}_c(\vec k + \vec q)\\
	&\displaystyle\times\frac{f(\epsilon_b(\vec k))-f(\epsilon_c(\vec k + \vec q))}{iq_m+\epsilon_b(\vec k)-\epsilon_c(\vec k+\vec q)}
	\end{array}
	\label{eq:nonintsuscep}
	\end{equation}
	where matrix elements $a_{\mu,a}$ as well as energy eigenvalues
	$\epsilon_a(\vec k)$ are obtained by  diagonalization of the tight-binding
	Hamiltonian $H_\text{kin}$.
	In order to facilitate the
	assignment of indices, we will follow the convention that greek letters denote
	site indices and latin letters denote band indices.
	Moreover, the susceptibility depends on the
	difference of the thermal occupation probability
	$f(x)=\frac{1}{1+e^{x/(k_BT)}}$, which follows from Fermi-Dirac statistics, and
	bosonic Matsubara frequencies $iq_m=2m\pi~T$.
	For $iq_m=0$, the denominator becomes divergent for equal band energies, which
	we treat by means of the rule of l'Hospital.
	Due to the fact that the considered Hamiltonians bear no non-local
	interactions, for susceptibilities of the form
	$\chi_{\mu\mu\nu\nu}$ (see below) we
	can  reduce the tensor product of vector spaces from
	$\mathbb{C}^N\otimes\mathbb{C}^N\otimes\mathbb{C}^N\otimes\mathbb{C}^N$ to
	$\mathbb{C}^{N}\otimes\mathbb{C}^{N}$.
	
	Spin and charge fluctuations within TPSC are treated by spin and charge
	susceptibilities ($\chi^\text{sp}$ and $\chi^\text{ch}$ respectively) from
	linear response theory.  \begin{equation}
	\begin{array}{rl}
	\chi^\text{sp}(\vec q, iq_m) &= [\mathbb{I}-U^\text{sp}\chi^{0}(\vec q,iq_m)]^{-1}2\chi^0(\vec q,iq_m)\\
	\chi^\text{ch}(\vec q, iq_m) &= [\mathbb{I}+U^\text{ch}\chi^{0}(\vec q,iq_m)]^{-1}2\chi^0(\vec q,iq_m).
	\end{array}
	\label{eq:chi_sp/ch}
	\end{equation}
	The renormalized irreducible vertices in the spin channel $U^\text{sp}$ and in the charge channel $U^\text{ch}$ are determined by local spin and charge sum rules
	\begin{equation}
	\begin{array}{rl}
	\frac{T}{N_{\vec q}}\sum\limits_{\vec q,iq_m}\chi^\text{sp}_{\mu\mu}(\vec q, iq_m) &= n_\mu - 2\langle n_{\mu\uparrow}n_{\mu\downarrow}\rangle\\
	\frac{T}{N_{\vec q}}\sum\limits_{\vec q,iq_m}\chi^\text{ch}_{\mu\mu}(\vec q, iq_m) &= n_\mu + 2\langle n_{\mu\uparrow}n_{\mu\downarrow}\rangle - n^2_\mu,
	\end{array}
	\label{eq:sumrules}
	\end{equation}
	where the spin vertex $U^\text{sp}$ is calculated from an ansatz equation that is motivated by the Kanamori-Brueckner screening~\cite{Vilk1997}
	\begin{equation}
	\begin{array}{rl}
	U^\text{sp}_{\mu\nu} = \frac{\langle n_{\mu\uparrow}n_{\mu\downarrow}\rangle}{\langle n_{\mu\uparrow}\rangle \langle n_{\mu\downarrow}\rangle}U\delta_{\mu,\nu}
	\end{array}
	\label{eq:Uspansatz}
	\end{equation}
	and the diagonal elements of the charge vertex $U^\text{ch}_{\mu\mu}$ are directly calculated from the local charge sum rule while off-diagonal elements are zero.
	Correlation effects within the Green's function $G$ are taken into account using a single-shot self-energy $\Sigma$ and incorporated by the Dyson equation
	\begin{widetext}
		\begin{align}
		\label{eq:GSigma}
		\Sigma_{\mu\nu}(\vec k,i\omega_n) &= Un_{\mu,-\sigma}\delta_{\mu,\nu}
		+\frac{UT}{8N_{\vec q}}\sum\limits_{\vec q, iq_m}\left[3U^\text{sp}_{\mu\mu}\chi^\text{sp}_{\mu\nu}(\vec q, iq_m)+U^\text{ch}_{\mu\mu}\chi^\text{ch}_{\mu\nu}(\vec q, iq_m)\right]G^0_{\mu\nu}(\vec k-\vec q, i\omega_{n-m})\\
		G_{\mu\nu}(\vec k, i\omega_n) &= \left[{G^0}^{-1}_{\mu\nu}(\vec k, i\omega_n)- \Sigma_{\mu\nu}(\vec k, i\omega_n)\right]^{-1}.
		\end{align}
	\end{widetext}
	In this framework, we employ Migdal-Eliashberg theory to calculate the
	superconducting gap $\Delta_{\mu\nu}(\vec k, i\omega_n)$.  We restrict our
	calculations to singlet and even-frequency and -orbital solutions, i.e.
	\begin{equation}
	\Delta_{\mu\nu}(\vec k, i\omega_n) = \Delta_{\mu\nu}(-\vec k,i\omega_n) = \Delta_{\mu\nu}(\vec k,-i\omega_n).
	\label{eq:gapsymmetry}
	\end{equation}
	The linearized Eliashberg equation takes the form
	\begin{widetext}
		\begin{align}
		\lambda\Delta_{\mu\nu}(\vec k,i\omega_n) &= \frac{T}{N_{\vec k'}}\sum_{\vec k',i\omega_{n'}}V_{\mu\nu}(\vec k-\vec k',iq_{n-n'})\sum_{\alpha,\beta}G_{\mu\alpha}(\vec k',i\omega_{n'})\Delta_{\alpha\beta}(\vec k',i\omega_{n'})G^\ast_{\nu\beta}(\vec k',i\omega_{n'}) \label{eq:Eliashberg}
		\end{align}
	\end{widetext}
	where the temperature at which the largest positive eigenvalue $\lambda$ becomes unity indicates the onset of superconductivity.
	The singlet pairing potential is calculated within the Random Phase Approximation (RPA~\cite{BickersScalapinoWhitePairingVertex,
		ScalapinoLohHirsch}) and given by
	
	\begin{eqnarray}
	V(\vec q,iq_m) = -\frac{3}{4}U^\text{sp}\chi^\text{sp}(\vec q, iq_m)U \nonumber
	+\frac{1}{4}U^\text{ch}\chi^\text{ch}(\vec q, iq_m)U - \frac{1}{2}U.
	\end{eqnarray}
	
	We enforce singlet solutions by symmetrization of the gap $(G\Delta G)_{\mu\nu}^s(\vec k,i\omega_n) =\frac{1}{2}\left[(G \Delta G)_{\mu\nu}(\vec k,i\omega_n)+(G \Delta G)_{\nu\mu}(-\vec k,-i\omega_n)\right]$ entering on the right-hand-side of the linearized Eliashberg equation (Eq. \ref{eq:Eliashberg}).
	For the numerical evaluation of the non-interacting susceptibility
	we employed adaptive cubature based on a three-point formula for triangles
	with an integration tolerance of $10^{-6}$. The interacting susceptibilities
	are strongly peaked when approaching the critical temperature. Therefore they
	were calculated on a $200\times200$ k grid for the molecule model and
	$300\times300$ k grid for the dimer model, while all other quantitities were
	well-converged on $70 \times 70$ grids.  For the evaluation of
	Eqs.~\ref{eq:GSigma} and~\ref{eq:Eliashberg}, we additionally employed fast
	Fourier transforms and the circular convolution theorem for a highly efficient
	implementation.
	The summation over Matsubara frequencies was performed for $N_\text{Mats}= 40\cdot(0.025/T)$ points, whereas high-frequency corrections up to the order of $\frac{1}{\omega^2}$ were included by extrapolation.
	
	\section{Results and Discussion}

	\subsection{Half-filled dimer model}
	Although the insufficiency of the dimer model for capturing the physics of the $\kappa$-(ET)$_2$X systems has been discussed~\cite{Guterding2016,Watanabe2017}, we will first use this well-explored model as a benchmark for our TPSC calculations.

\begin{figure}[t]
		\includegraphics[width=0.48\linewidth]{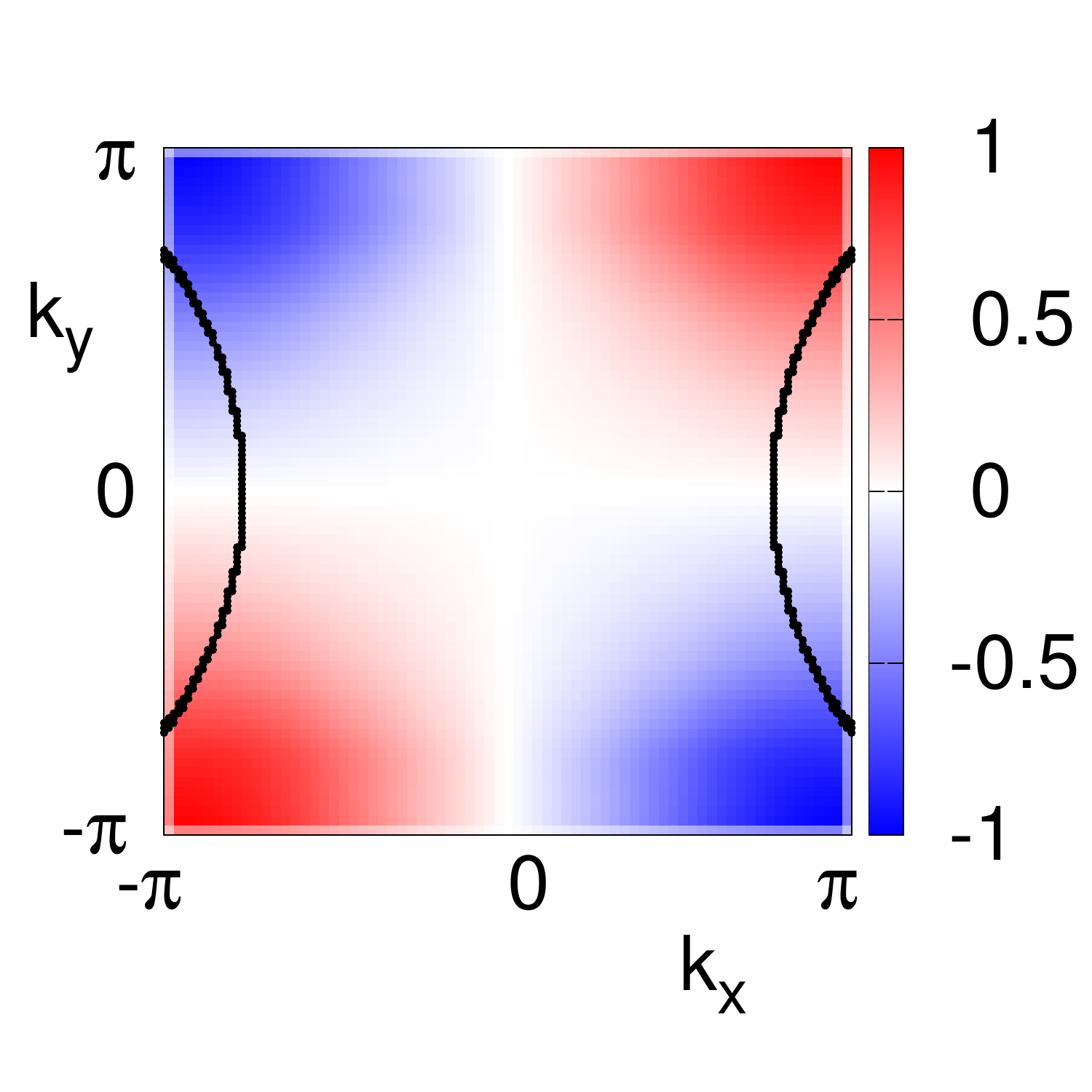}
		\includegraphics[width=0.48\linewidth]{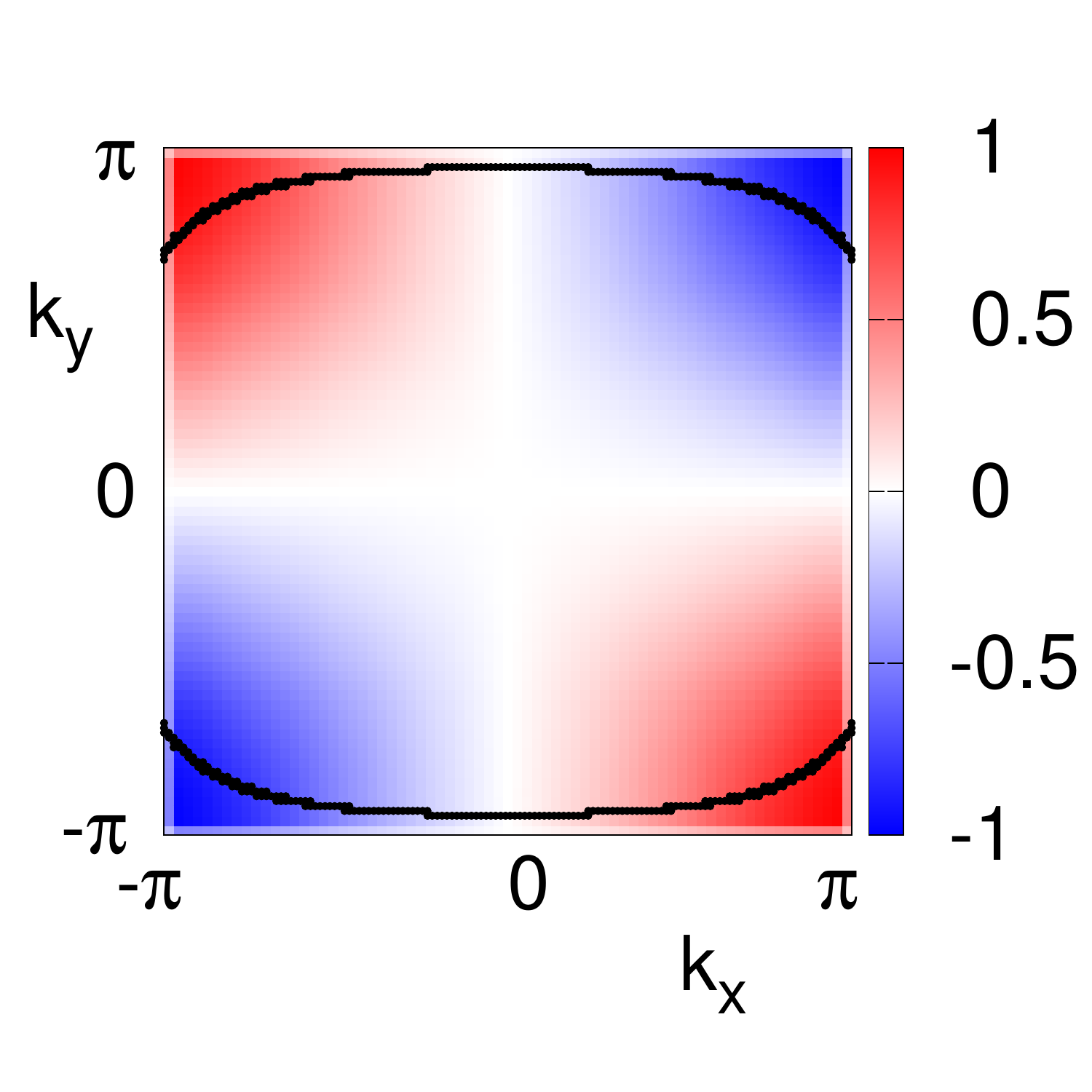}
		\caption{Superconducting gap $\Delta(\vec k,i\omega_0)$ of
			the dimer model in the first physical Brillouin zone (see main text).
			The dominant d$_{xy}$ character shows nodes along the boundaries, since it has to be 2$\pi$ periodic, and a sign change between (a) the first band and (b) the second band that has been previously assigned to strong inter-band coupling~\cite{Schmalian1998}.}
		\label{fig:dimer_symmetry}
	\end{figure}

	In the context of the high-Tc cuprate superconductors it is already well known that the half-filled single-band  Hubbard model on the square lattice stabilizes d$_{x^2-y^2}$ pairing solutions. Introducing anisotropic diagonal couplings $t'$ one expects superconductivity to
	become unstable for high values of the frustration $t'/t$, while the d-wave solution will be retained for intermediate frustration strengths, as it is the case for the half-filled single-band triangular lattice model for $\kappa$-(ET)$_2$X.
	In order to compare the results of this section to the molecule model considered in the following section, we have to transform this solution to the physical Brillouin zone (BZ) of the $\kappa$-(ET)$_2$X (corresponding to two dimers per crystallographic unit cell), which is half as large as of the single-band model. 
	Folding the BZ corners and rotating by 45\degree ~\cite{Guterding2016}, we expect a d$_{xy}$ solution with gap maxima at ($\pm \pi,\pm \pi$).
	Additionally, the $2\pi$ periodicity of the gap function enforces node-lines
	along the BZ boundaries ($k_x,\pi$) and ($\pi,k_y$). Note that there is a sign
	change between the two bands at the Fermi surface (see
	Fig.~\ref{fig:dimer_symmetry}), which previously  has been attributed to strong
	inter-band coupling~\cite{Schmalian1998}. 
	
	\begin{figure}[t]
		\includegraphics[width=\linewidth]{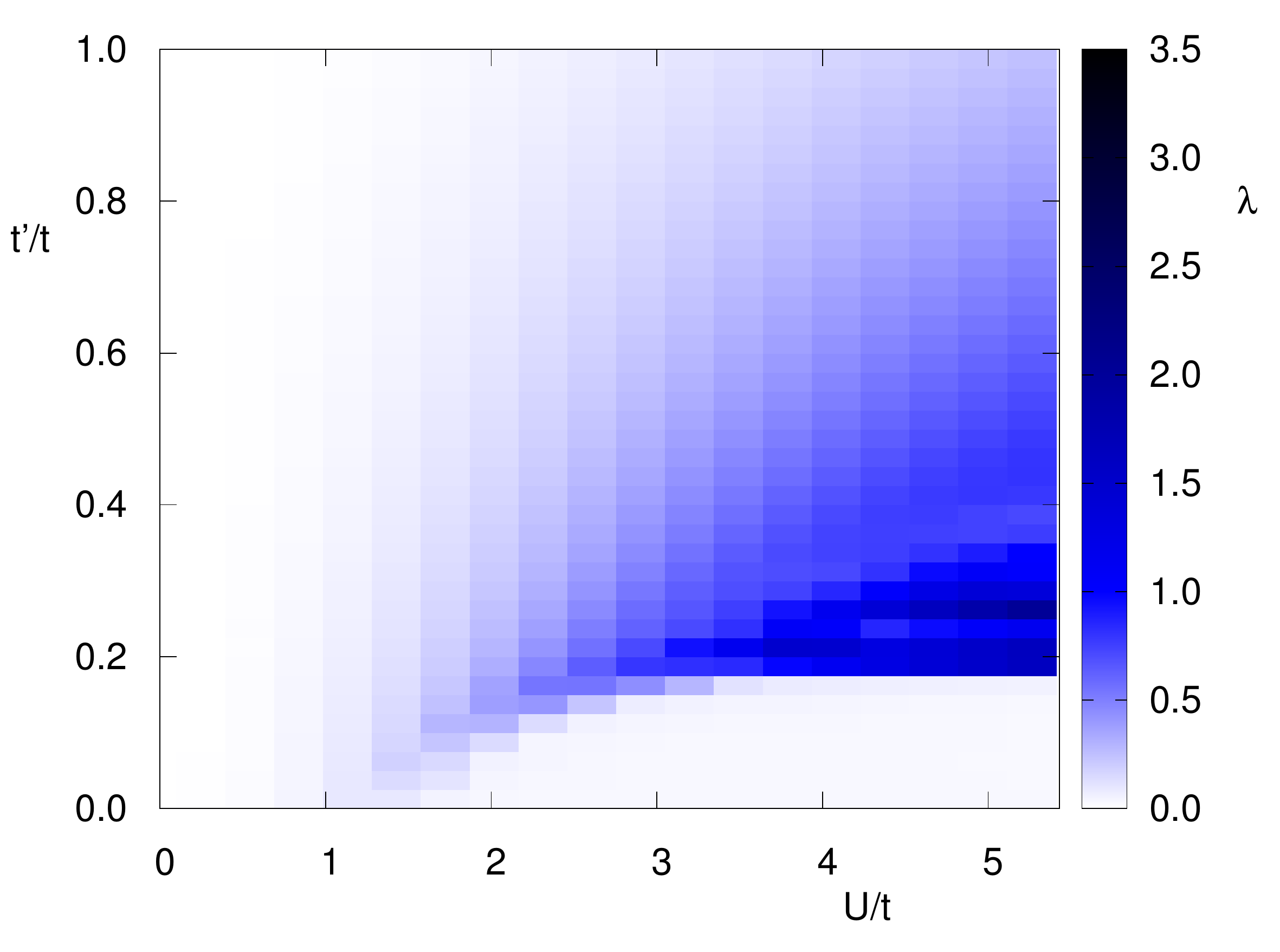}
		\caption{Largest positive eigenvalue of the linearized Eliashberg equation at T=0.003~eV within the dimer model. Moderate on-site interactions are crucial to obtain superconductivity, while too strong correlations ($U_\text{dim}/t\gtrsim1.5$) result in the opening of a pseudogap and Mott insulator physics.  Strong afm insulating tendencies
			can be reduced by next nearest neighbor hoppings and the implicit geometric
			frustration. A combination of both ($t'/t\approx0.25$ and $U_\text{dim}/t>3.4$)
			is most favorable for superconductivity.}
		\label{fig:dimer_phasediagram}
	\end{figure}
	
	\begin{table*}[t] 
		\setlength{\tabcolsep}{3mm}
		
		\begin{tabular}{l||ll|l|ll}
			material & $t'/t$ & $t_4/t_2$ & $U_\text{dim}$ [eV] & $T^\text{TPSC}_c$ [K] & $T_c$ \cite{Hiramatsu2015,Kato1987,Mori1990,Kini1990} [K]\\
			\hline \hline
			$\kappa$-(ET)$_2$Ag(CF$_3$)$_4$(TCE) & 0.449 & 0.362 & 0.336 & 32.5 & 2.6\\
			$\kappa$-(ET)$_2$I$_3$ & 0.346 & 0.266 & 0.36 & 44.1 & 3.6\\
			$\kappa$-(ET)$_2$Ag(CN)$_2$I$\cdot$H$_2$O & 0.473 & 0.305 & 0.37 & 31.3 & 5.0\\
			$\kappa$-$\alpha'_1$-(ET)$_2$Ag(CF$_3$)$_4$(TCE) & 0.495 & 0.362 & 0.332 & 27.1 & 9.5\\
			$\kappa$-(ET)$_2$Cu(NCS)$_2$ & 0.69 & 0.171 & 0.38 & 15 & 10.4\\
			$\kappa$-$\alpha'_2$-(ET)$_2$Ag(CF$_3$)$_4$(TCE) & 0.495 & 0.369 & 0.33 & 26.2 & 11.1\\
			$\kappa$-(ET)$_2$Cu[N(CN)$_2$](CN) & 0.669 & 0.172 & 0.35 & 13.9 & 11.2\\
			$\kappa$-(ET)$_2$Cu[N(CN)$_2$]Br & 0.455 & 0.379 & 0.354 & 32.5 & 11.6
		\end{tabular}
		\caption{Comparison of the calculated and experimental critical temperatures $T_c$ for several organic charge transfer salts. The calculations within the dimer model do not reproduce the general trend of the experimental results but can be understood by means of geometric frustration (see Fig.~\ref{fig:Tc-tt-plot}).}
		\label{tab:Tc_dimer}
	\end{table*}

	\begin{figure}[t]
		\includegraphics[width=\linewidth]{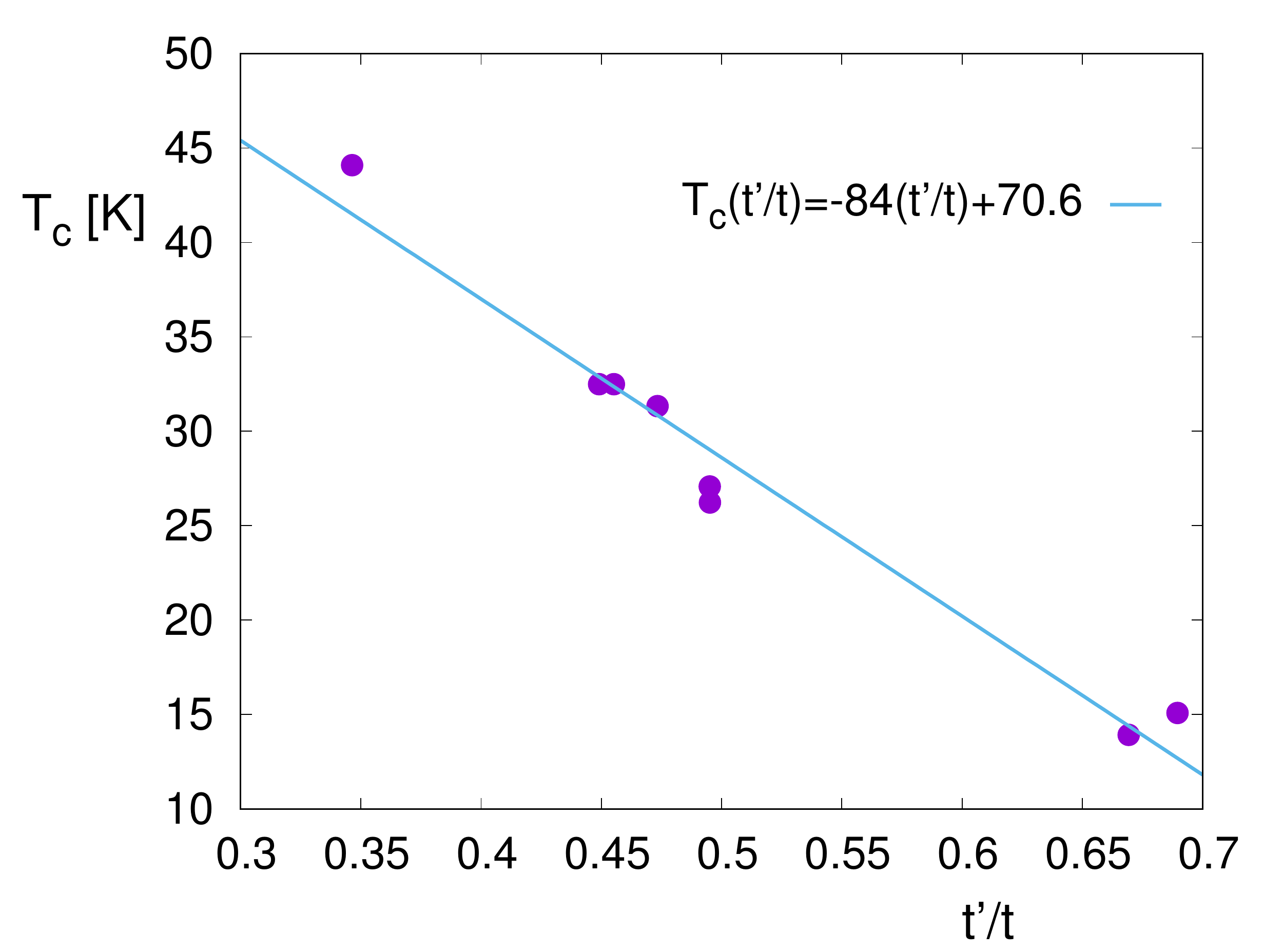}
		\caption{
			Critical temperature T$_c$ calculated within the combined TPSC + Eliashberg approach as a function of $t'/t$ for the
			eight $\kappa$-(ET)$_2$X materials listed in Table~\ref{tab:Tc_dimer}. T$_c$ drops monotonically with increasing $t'/t$ since the 
			geometric
			frustration suppresses an afm state und reduces therefore spin
			fluctuations that are key ingredients for large T$_c$.
			As a guide to the eye we show the linear fit to the data points.} \label{fig:Tc-tt-plot}
	\end{figure}
	
	In order to explore the role of the diagonal hopping $t'$ we have calculated the largest positive eigenvalue of the linearized Eliashberg equation (Eq. \ref{eq:Eliashberg}) at $T=0.003$~eV$\approx 35$~K in dependence of the hopping ratio $t'/t$ and the relative on-site repulsion $U_\text{dim}/t$ (Fig.~\ref{fig:dimer_phasediagram}), where a large eigenvalue implies a close proximity to the superconducting state that is realized at $\lambda=1$.
	We find that several effects compete: at large $t'/t$ ratios the
	antiferromagnetic (afm) fluctuations that drive  superconductivity, are
	strongly suppressed, while for large correlations a pseudogap opens, reducing
	the number of states close to the Fermi level and therefore the total energy
	gained by the formation of superconducting pairs.  Therefore, 
	both effects are needed to enhance  spin fluctuations but they should
	be kept moderate
	enough to prevent magnetic ordering.

	Finally, we calculate critical temperatures for eight representatives of the
	$\kappa$-(ET)$_2$X family for $U_\text{dim}\approx2t_1$ (see Tab.~\ref{tab:Tc_dimer}). We
	observe no
	obvious relation to the measured critical temperatures. 
	However, plotting the calculated
	critical temperatures against the corresponding frustration values
	(Fig.~\ref{fig:Tc-tt-plot}), we find a monotonous decrease with increasing
	geometric frustration, i.e., the diagonal hopping suppresses the afm spin
	fluctuations that drive the superconductivity. As the measured critical
	temperatures do not follow this simple trend, it is obvious that we have to go
	beyond the dimer model in order to understand the superconductivity in the
	$\kappa$-(ET)$_2$X family.

	\subsection{3/4-filled molecule model}
	\begin{figure}[t]
		\includegraphics[width=0.48\linewidth]{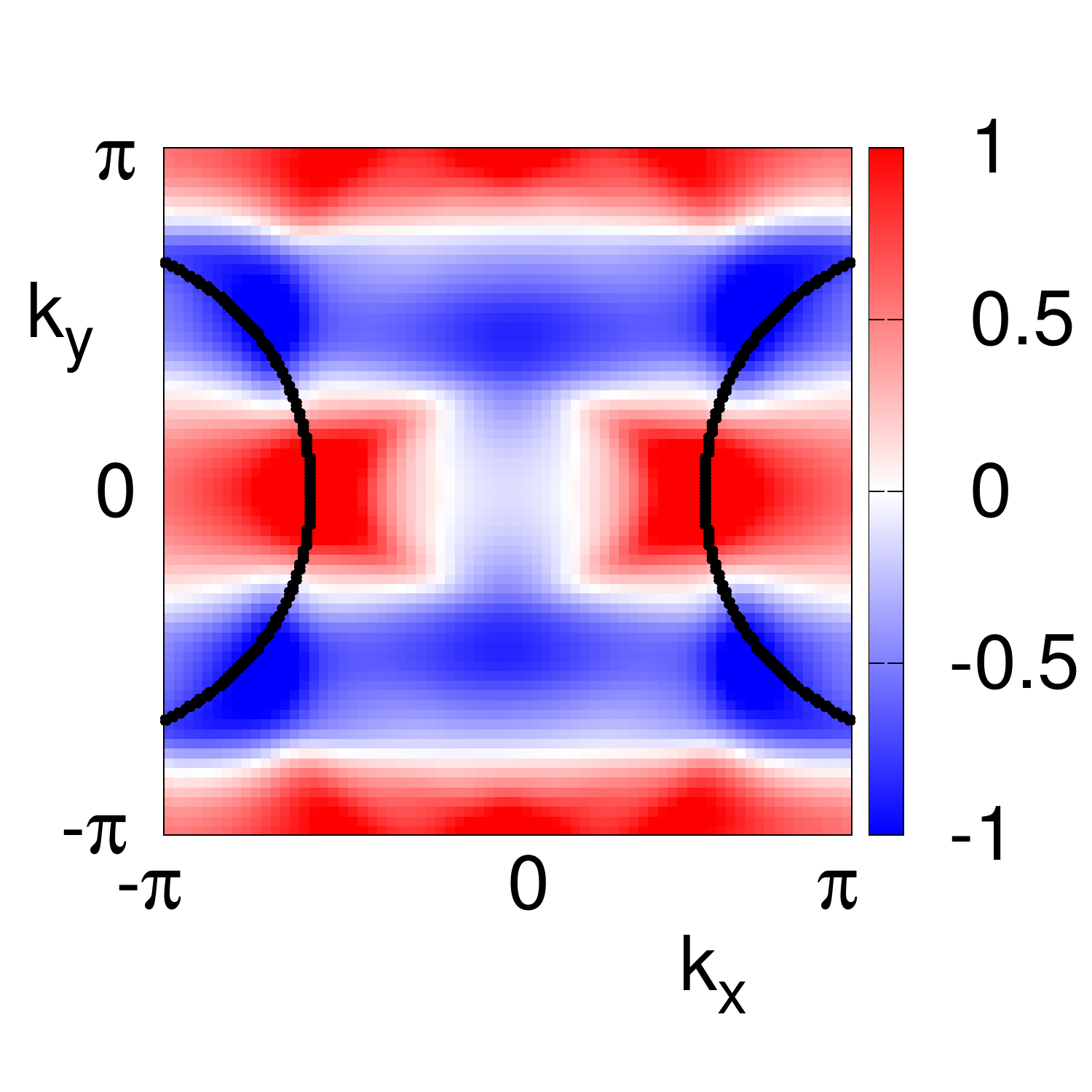}
		\includegraphics[width=0.48\linewidth]{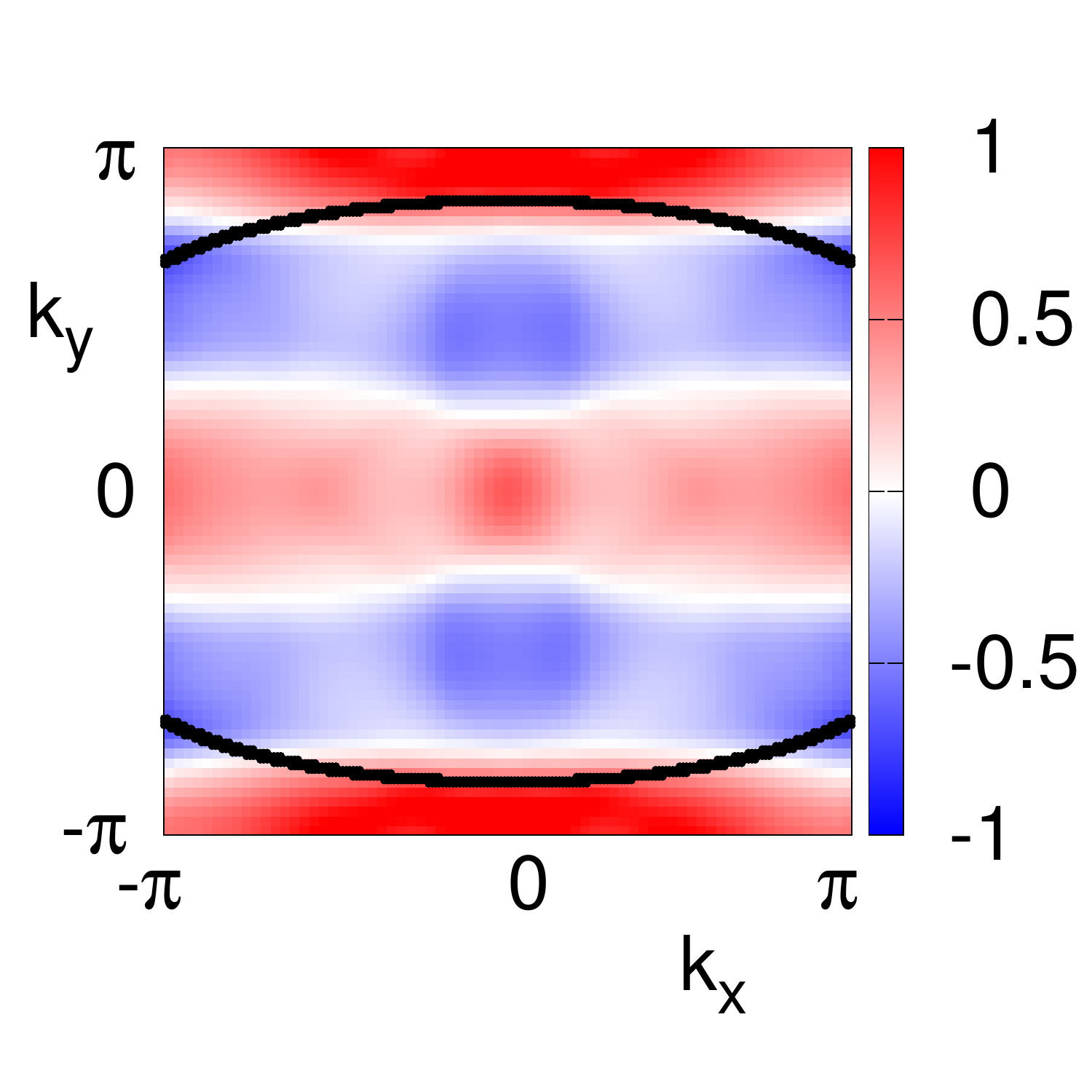}
		\caption{s$_{\pm}$+d$_{x^2-y^2}$ superconducting gap $\Delta(\vec k,i\omega_0)$ as obtained in the four band molecule model at low temperatures. This kind of symmetry was already observed in Ref.~\onlinecite{Guterding2016} in the context of RPA calculations.}
		\label{fig:s_pm_dx2y2_gap}
	\end{figure}

	\begin{figure}[t]
		\includegraphics[width=\linewidth]{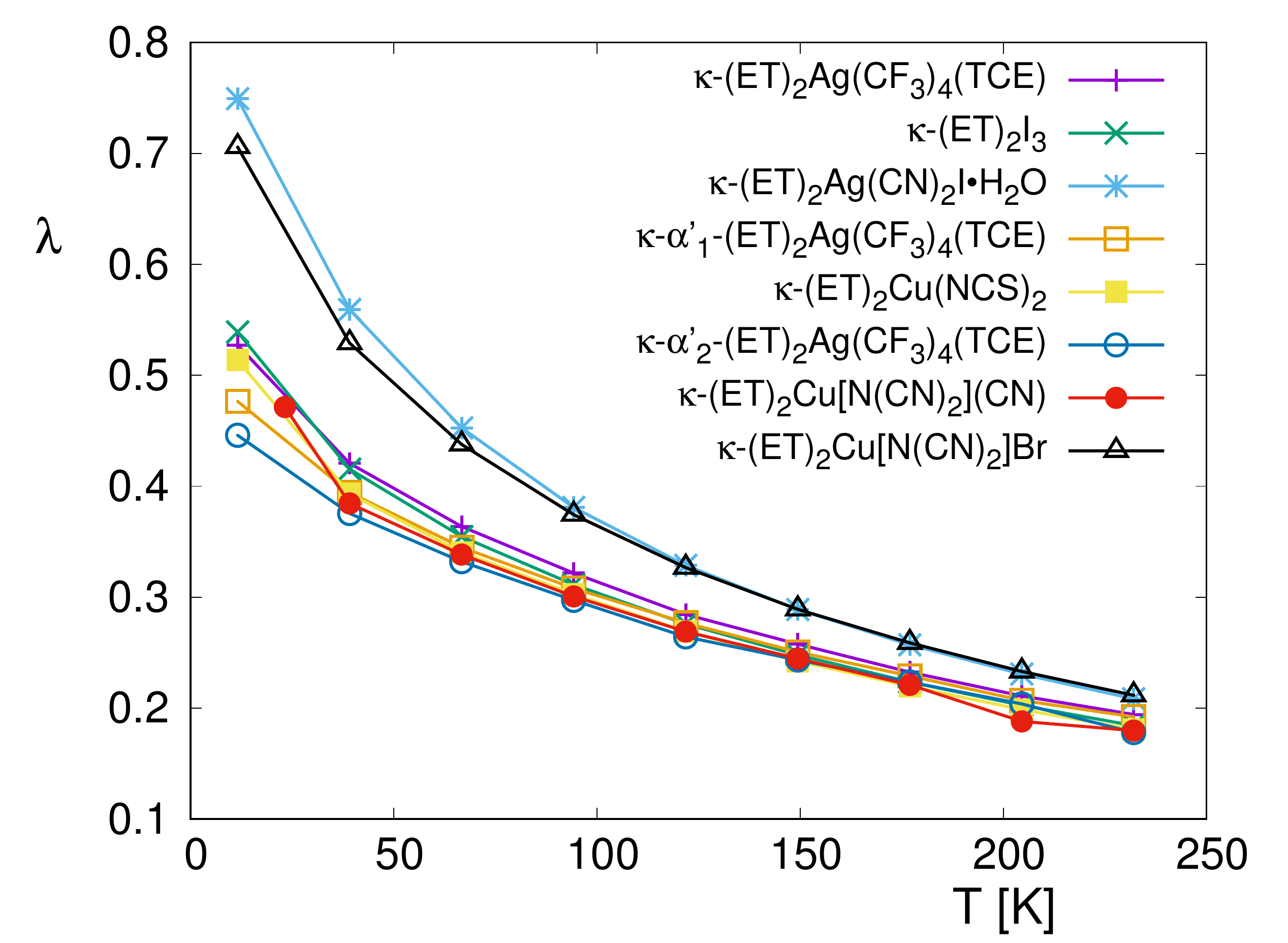}
		\caption{Largest positive eigenvalue $\lambda$  of the linearized Eliashberg equation for $U_\text{mol}=0.65$~eV. We see only small differences between the values for each material which can be understood from the similarity of the hopping parameters. The four band model based on the largest hopping elements is not sufficient to reproduce the trends of $T_c$.}
		\label{fig:lambda-T-plot}
	\end{figure}

	\begin{figure}[t]
		\includegraphics[width=\linewidth]{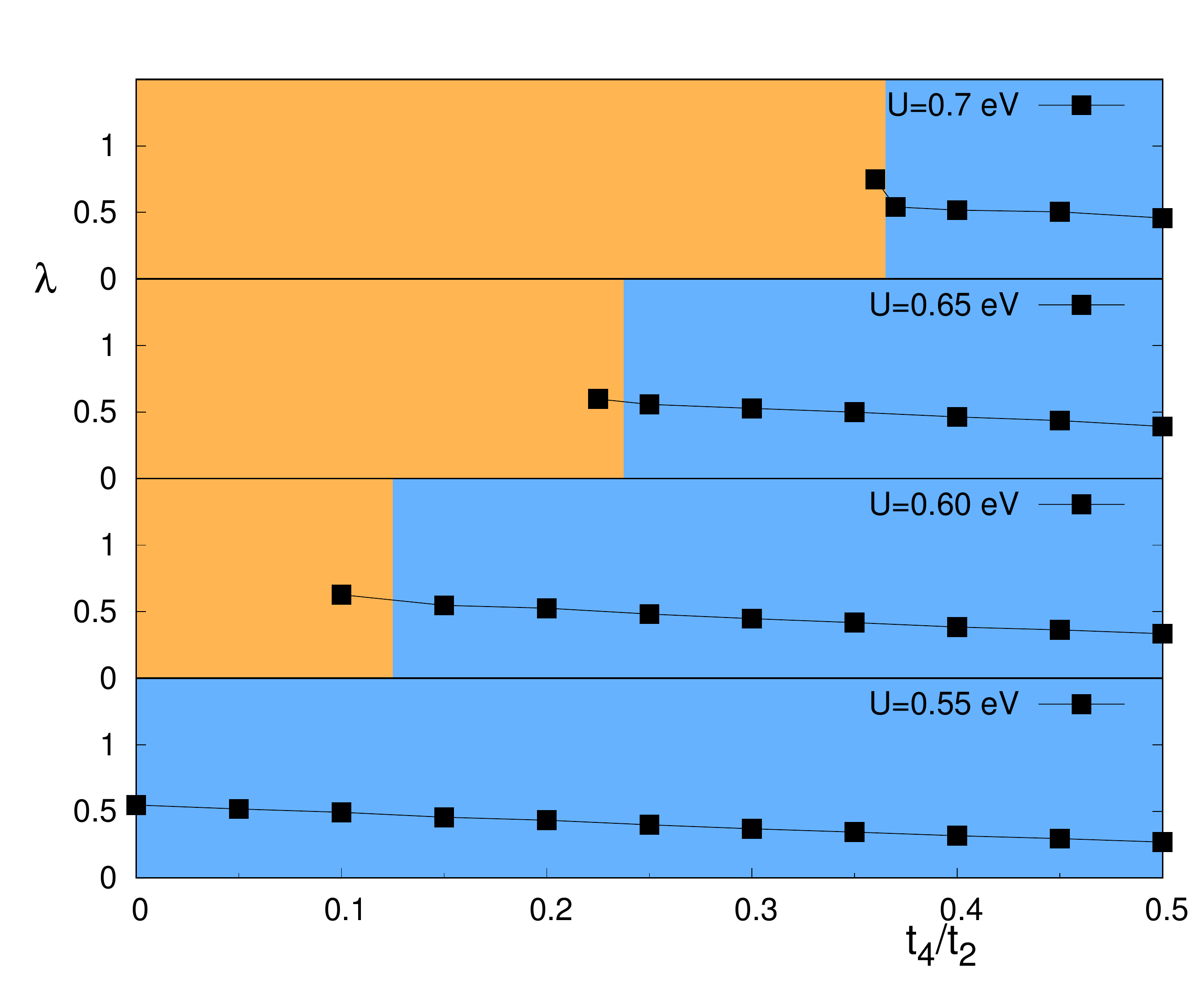}
		\caption{Largest positive eigenvalue $\lambda$  of the linearized Eliashberg equation for $U_\text{mol}=0.55, 0.6, 0.65\text{ and } 0.7~eV$ ($t_1,t_2$ and $t_3$ averaged from GGA results~\cite{Guterding2016}). We see an overall increase in $\lambda$ by going to larger values of $U_\text{mol}$ and an inset of extended $s+d_{x^2-y^2}$ gap symmetry (orange background) at small values of $t_4/t_2$ while $d_{xy}$ symmetry (blue background) is dominant otherwise.}
		\label{fig:lambda-t4t2-plot}
	\end{figure}
	We consider now the 
	proposed four-parameter molecule model~\cite{Guterding2016}
	as the starting point of the present
	study (the comparison to the full \textit{ab initio} derived tight-binding
	model is discussed in Appendix~\ref{sec:susc_symm}).
	
	In the molecule model, the center of each ET molecule constitutes a lattice site in the tight binding model yielding a three quarter filled four-band system
	(four molecules per unit cell).
	Compared to the dimer model two additional degrees of freedom are accessible: the strength of the intra-dimer hopping and the in-plane anisotropy.
	In a previous study~\cite{Guterding2016} it was demonstrated that already a small degree of in-plane anisotropy results in considerable symmetry changes of the gap function. 
	For all realistic parameter sets the exotic eight-node gap function was found to be favorable, whereas several compounds were shown to lie close to a phase boundary to d$_{xy}$ symmetry. 
	
	As the previous static RPA approach may only give qualitative results on the gap symmetry, in this study we apply the above introduced combination of the linearized Eliashberg equation, the RPA expression for the pairing vertex and the TPSC self-energy corrected Green's functions and renormalized vertices.
	
	Interestingly,  the gap symmetries calculated with the more advanced
	TPSC approach differ from the RPA predictions. For moderate on-site molecule $U_\text{mol}$ values, only 
	$\kappa$-(ET)$_2$Cu(NCS)$_2$ and $\kappa$-(ET)$_2$Cu[N(CN)$_2$](CN), with the highest in-plane anisotropies feature s$_{\pm}$+d$_{x^2-y^2}$ symmetry (see Fig.~\ref{fig:s_pm_dx2y2_gap}), while the other compounds exhibit a simple d$_{xy}$ symmetry as in the dimer model. Only for larger values of $U_\text{mol}$ s$_{\pm}$+d$_{x^2-y^2}$ is stabilized in a large region in parameter space including all superconducting materials studied with RPA if $U_\text{mol}\gtrsim0.7eV$ (see Fig. 7). 
	This result and the experimental evidence of s$_\pm$+d$_{x^2-y^2}$ gap
	symmetry in $\kappa$-(ET)$_2$Cu[N(CN)$_2$]Br~\cite{Diehl2016} indicate the
	importance of correlations in these materials not only for the enhancement of
	spin-fluctuations but also for the symmetry of the gap function. Although 
	the TPSC approach - in general - also allows us to determine the critical temperatures
	for superconductivity, we find that correlation effects give rise to strong antiferromagnetic fluctuations 
(as indicated by diverging spin susceptibilities) in TPSC, that do not 
allow us to obtain meaningful results in the $\lambda \approx 1$ regime
for the molecule model.\footnote{Although it is possible to obtain $\lambda\geq
		1$ below $\sim 10$~K, those solutions are not reliable since the corresponding
		susceptibilities are strongly peaked and the numerical integrations yield
		increasingly large errors.} Therefore, we can only estimate trends for the
	critical temperatures from the magnitude of the eigenvalue of the Eliashberg
	equation at higher temperatures above the superconducting transition, as
	displayed in Fig.~\ref{fig:lambda-T-plot}, where we assumed
	$U_\text{mol}=0.65~$eV\footnote{For higher $U_\text{mol}$ values it was not
		possible to perform calculations down to temperatures $T\approx10K$ since the
		spin susceptibilities of some compound start to diverge and hamper numerical
		integrations. However our calculations (Fig.~\ref{fig:lambda-t4t2-plot}) show
		that the parameter region for extended $s+d_{x^2-y^2}$ gap symmetry increases
		with larger correlation strenghts.}.
	We find that although the two materials with the  $s_{\pm}+d_{x^2-y^2}$ solution show the strongest deviations in the hopping parameters, they are located on the same branch as most of the other materials.
	Instead,  $\kappa$-(ET)$_2$Ag(CN)$_2$I$\cdot$H$_2$O and $\kappa$-(ET)$_2$Cu[N(CN)$_2$]Br, 
	exhibit especially high eigenvalues in the considered temperature range that do not coincide with experimental observations.  Nevertheless, it is interesting to note that also the $s_{\pm}+d_{x^2-y^2}$ compounds prefer a $d_{xy}$ solution at high temperatures above the superconducting transition (see appendix~\ref{sec:gap_T}). 
	Only below $\sim 20$~K the eight-node solution is stabilized, although the
	susceptibilities already reveal the tendency towards the  $s_{\pm}+d_{x^2-y^2}$
	solution at higher temperatures (see Fig.~\ref{fig:5_eigenvalue} in the appendix~\ref{sec:gap_T}).
	While the spin susceptibilities of the materials displaying $d_{xy}$ symmetry in the considered parameter range peak at reciprocal
	vectors $q\sim(0.6\pi,0.37\pi)$, the peaks are significantly shifted to higher
	$q_x$ values for the two compounds with high in-plane anisotropies,
	$q\sim(0.76\pi,0.41\pi)$. 
	Moreover, as mentioned above,
	in the TPSC  calculations  the magnitude of the
	on-site Hubbard repulsion strongly influences the gap symmetry. 
	At low values of the Hubbard interaction it is not possible to access the $s_{\pm}+d_{x^2-y^2}$ region for any strength of the in-plane anisotropy, while slightly larger values shift the transition line towards $t_4/t_2$ values of up to $0.365$ for $U_\text{mol}=0.7eV$ (see Fig.~\ref{fig:lambda-t4t2-plot}).

	
	\section{Summary and Outlook}
	
	To summarize, we have applied a combination of TPSC and the Eliashberg framework for superconductivity in order to derive the gap symmetries and trends for the critical temperatures in the dimer and molecule models for several superconducting $\kappa$-(ET)$_2$X materials. 
	Within the dimer model we find that the critical temperatures only reflect the
	frustration of the system but do not reproduce the experimental trends. 
	Our calculations for the molecule model confirm  previous findings that the
	additional degrees of freedom, i.e. the intra-dimer hopping and the in-plane
	anisotropy, are decisive for the gap symmetry and can result in $s_{\pm}+d_{x^2-y^2}$ solutions. However, we find that the $s_{\pm}+d_{x^2-y^2}$ gap is further  stabilized by increasing correlations and may therefore be realized within the range of realistic model parameters.
	These three tuning parameters are known to be very sensitive to pressure or strain as well as to endgroup disorder.
	Switching between the different gap symmetries may therefore be easily realizable and should be observable in f.i. state-of-the-art scanning tunneling spectroscopy measurements.

	\begin{acknowledgments}
		This work was supported by the 
		German Research Foundation (Deutsche Forschungsgemeinschaft) under grant SFB/TR 
		49. Calculations were performed on the LOEWE-CSC supercomputers of the 
		Center for Scientific Computing (CSC) in Frankfurt am Main, Germany.
	\end{acknowledgments}
	
	\appendix
	\clearpage
	\newpage
	\section{Temperature dependence of gap symmetry}
	\label{sec:gap_T}
\begin{figure}[tb]
		\includegraphics[width=\linewidth]{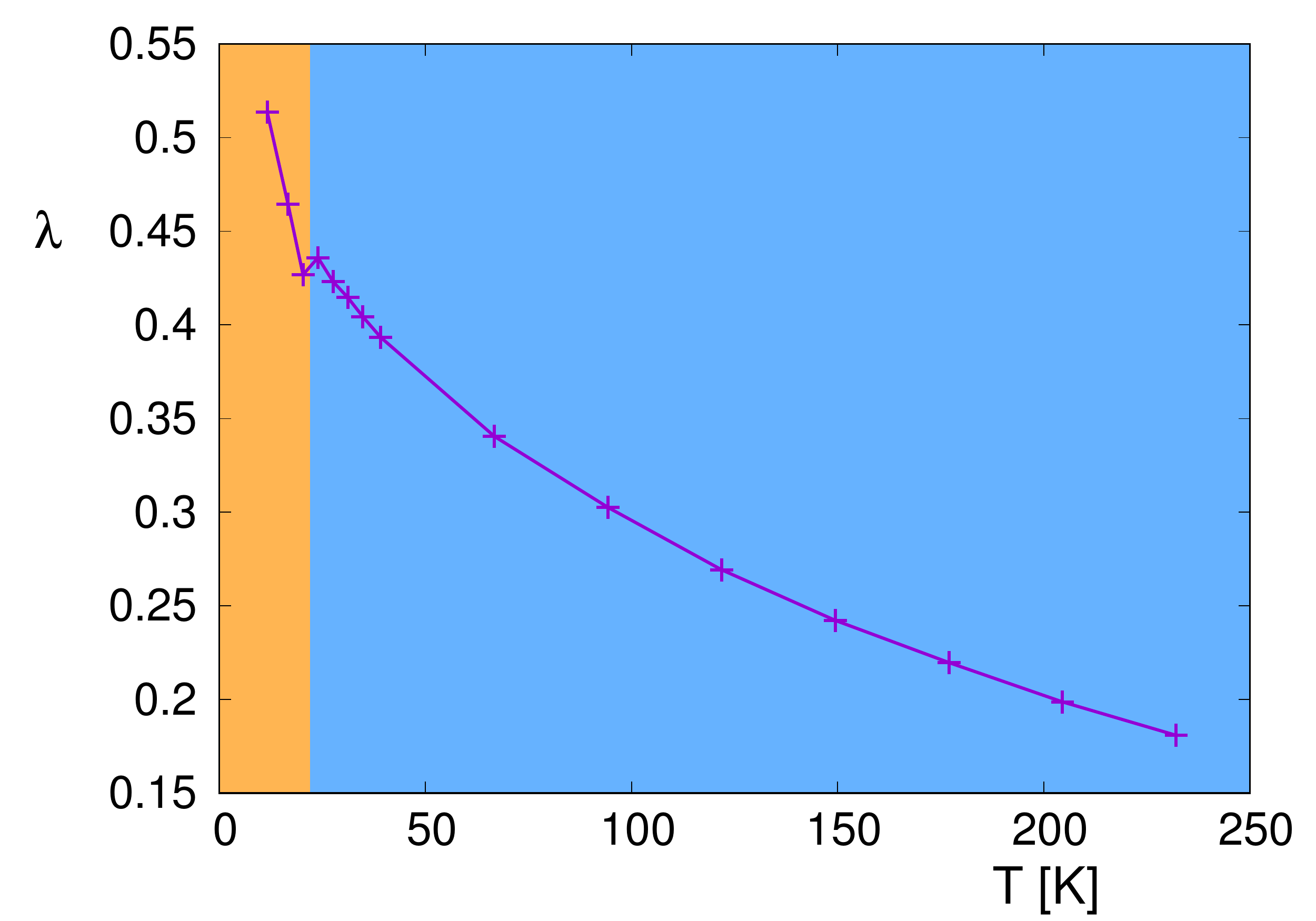}
		\caption{Largest eigenvalue $\lambda$ of the linearized Eliashberg equation for $\kappa$-(ET)$_2$Cu(NCS)$_2$ in the four-parameter model with $U_{mol}=0.65$~eV. A change in the gap symmetry (from d$_{xy}$ to extended s+d$_{x^2-y^2}$) becomes visible at T$\approx$22~K and is accompanied by a sudden drop in the eigenvalue that can be understood as a suppression of the superconducting state due to competing symmetries.}
		\label{fig:5_eigenvalue}
	\end{figure}

	Our combined TPSC and Eliashberg framework allows us to track the gap symmetry and the corresponding eigenvalue at temperatures above the superconducting transition. Interestingly, we find that at high temperatures, $T\gg T_c$, all materials yield a $d_{xy}$ symmetry. Only at temperatures close to the superconducting transition, the materials with high in-plane anisotropy and/or large correlations undergo a transition to the  extended s+d$_{x^2-y^2}$ symmetry accompanied by a change of the slope and a non-monotonous jump in the Eliashberg eigenvalue, as shown in Fig.~\ref{fig:5_eigenvalue} for the $\kappa$-(ET)$_2$Cu(NCS)$_2$ compound. This small drop in the eigenvalue can be interpreted in terms of a competition between the different order parameter symmetries, which destabilizes the superconducting state.
	
	\section{Susceptibilities and gap symmetries}
	\label{sec:susc_symm}
\begin{figure*}[htb]
		\includegraphics[width=0.8\linewidth]{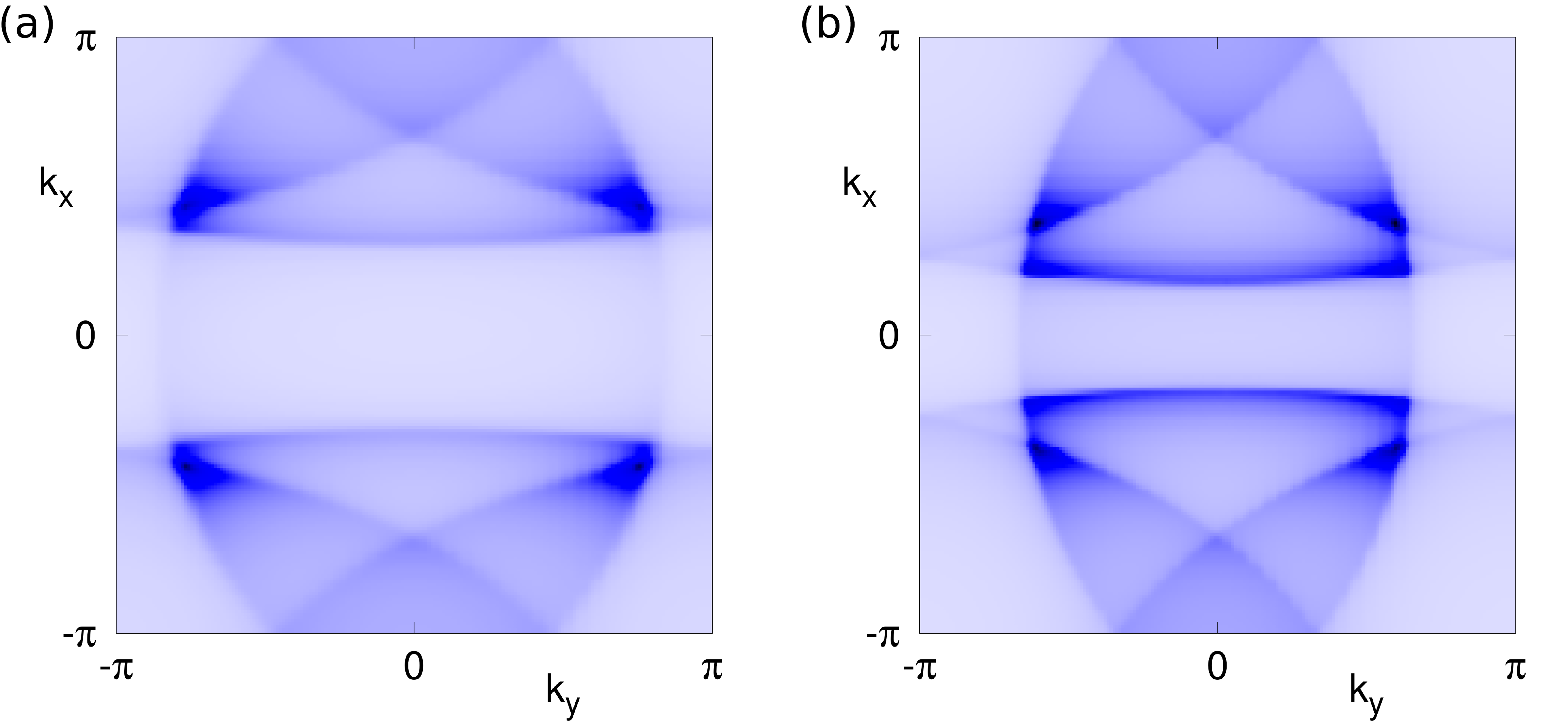}
		\caption{Static spin susceptibility $\sum_{\mu} \chi^{sp}_{\mu\mu}(\vec q, 0)$ for $s_{\pm}+d_{x^2-y^2}$ (a) and $d_{xy}$ (b) gap symmetries. The dominant spin fluctuations in the compounds with smaller in-plane anisotropy have wave vectors of about $(0.6\pi,0.37\pi)$, while the dominant vector is shifted towards large k values in the materials with high in-plane anisotropies,  $Q\sim(0.76\pi,0.41\pi)$.}
		\label{fig:peak-position}
	\end{figure*}

	Based on the discussion of the temperature dependence of the gap symmetries, we think that all materials might exhibit $s_{\pm}+d_{x^2-y^2}$ gap symmetry at very low temperatures, at which we can not perform meaningful calculations due to diverging factors in the linearized Eliashberg equation. In order to resolve this issue, we further investigate the driving force of the superconducting transition, i.e., the spin susceptibilities.
	Indeed, we can find two clear distinctions between the high in-plane anisotropy materials and the other $\kappa$-ET materials (see Fig.~\ref{fig:peak-position}): First, the broad shoulder connecting the one-dimensional parts of the Fermi surface is much less pronounced in the materials with high in-plane anisotropy. Second, the peak position is further shifted towards the Brillouin zone boundaries. While the strongest peaks in most of the materials are located at $(0.6\pi,0.37\pi)$, it is shifted to higher $k_x$, $k_y$ values, $(0.76\pi,0.41\pi)$, in $\kappa$-(ET)$_2$Cu(NCS)$_2$ and $\kappa$-(ET)$_2$Cu[N(CN)$_2$](CN).
	
	Hence, a careful inspection of the spin susceptibilities even at high temperatures can give clues as to the necessary strength of the correlations to realize  $s_{\pm}+d_{x^2-y^2}$ gap symmetries.
	
	\section{Reduction to largest hopping elements}
\begin{figure}[h]
		\includegraphics[width=\linewidth]{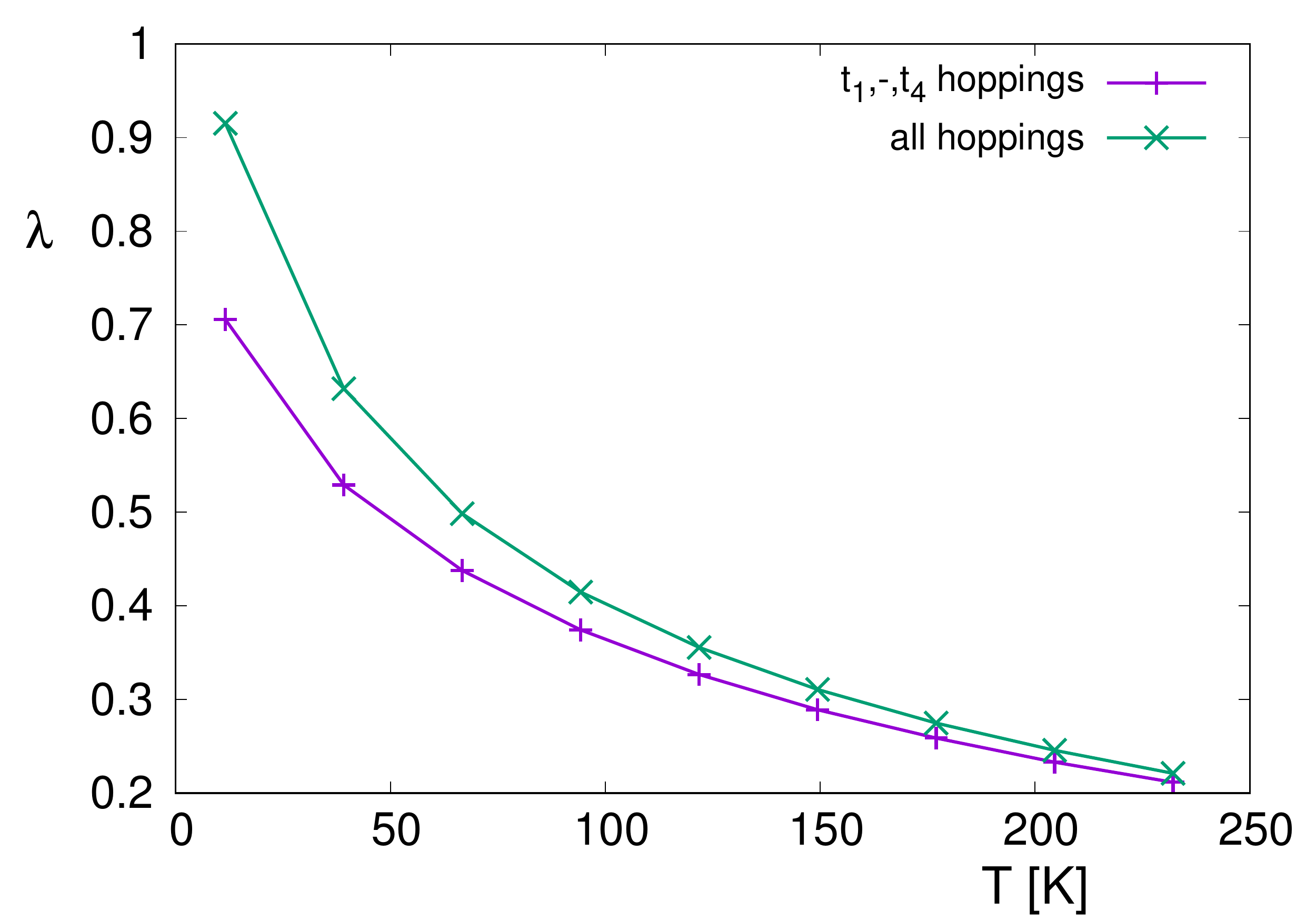}
		\caption{Temperature evolution of the largest Eliashberg eigenvalue $\lambda$ for the four-parameter (violet line) and full \textit{ab initio} derived kinetic Hamiltonian (green line) for $\kappa$-(ET)$_2$Cu[N(CN)$_2$]Br with $U_{mol}=0.65$~eV. While the gap symmetry remains unchanged, the inclusion of long-range hoppings stabilizes the superconducting state.}
		\label{fig:lambda_compare}
	\end{figure}

	In order to rule out the possibility that inclusion of further hopping parameters crucially influences the gap symmetries or transition temperatures, we have performed test calculation, where we compare the four-parameter calculations with the results of calculations, in which we take into account the full Hamiltonian as obtained by the Wannier projection method in DFT.  
	In Fig.~\ref{fig:lambda_compare} we show the temperature evolution of the largest Eliashberg eigenvalue for the two models for one representative of the $\kappa$-ET family. While we always obtain the same gap symmetry independent of the considered model, the eigenvalue is considerably enhanced due to the additional hoppings.
	Hence, we want to stress that accurate calculations of critical temperatures do not only have to carefully choose the strength of the Hubbard repulsion but also have to go beyond the four-parameter model.
	
	\bibliographystyle{apsrev4-1}
	\bibliography{Literatur.bib}
	
\end{document}